\documentclass[useAMS,usenatbib]{mn2e}
\usepackage{myaasmacros}
\usepackage{amsmath}
\usepackage{graphicx}
\usepackage{url}

\newcommand{\tabref}[1] {Table~\ref{#1}}
\newcommand{\figref}[1] {Figure~\ref{#1}}
\newcommand{\eqnref}[1] {Equation~\ref{#1}}
\newcommand{\secref}[1] {Section~\ref{#1}}
\newcommand{\bsec}{\farcs}

\newcommand\plotone[1]{%
 \centering
 \leavevmode
 \includegraphics[width={\columnwidth}]{#1}%
}%
\newcommand\plottwo[2]{{%
 \centering
 \leavevmode
 \columnwidth=.45\columnwidth
 \includegraphics[width={\columnwidth}]{#1}%
 \hfil
 \includegraphics[width={\columnwidth}]{#2}%
}}%

\title{Weak Microlensing}
\author[Coles et al.]{Jonathan Coles$^{1}$\thanks{E-mail: jonathan@physik.uzh.ch},
Prasenjit Saha$^{1}$, and
Hans Martin Schmid$^{2}$\\
$^{1}$Institut f\"ur Theoretische Physik, Universit\"at Z\"urich, 
Winterthurerstr. 190, 8006 Z\"urich, Switzerland\\
$^{2}$Institut f\"ur Astronomie, ETH Z\"urich, 8093 Z\"urich, Switzerland}
\begin{document}

\maketitle

\label{firstpage}

\begin{abstract}
A nearby star having a near-transit of a galaxy will cause a time-dependent
weak lensing of the galaxy. Because the effect is small, we refer to this as
weak microlensing. This could provide a useful method to weigh low-mass stars
and brown dwarfs.  We examine the feasibility of measuring masses in this way
and we find that a star causes measurable weak microlensing in a galaxy even at
10 Einstein radii away.  Of order one magnitude $I\leq25$ galaxy comes close
enough to one or other of the $\sim100$ nearest stars per year.  
\end{abstract}


\section{Introduction} \label{sec:Introduction}

Gravitational lensing by Galactic stars has come a long way since the
low-probability assessment of \cite{1936Sci....84..506E}.  The first detections
\citep{1993Natur.365..621A,1993Natur.365..623A,1993AcA....43..289U} have been
followed by $\sim4000$ more, including a few detections of planets around the
main lensing star \citep[e.g.,][]{2006ApJ...644L..37G}.

Yet despite the now abundant examples of microlensing, accurate measurements of
the lensing mass are still rare.  The reason is that a microlensing light-curve
on its own tells us precisely how long proper motion takes to traverse an
Einstein radius, but the physical length $r_E$ of the Einstein radius and its
angular size $\theta_E$ remain unknown.  Hence, only a broad statistical
statement about the stellar mass can be made in most cases.

\cite{1966MNRAS.134..315R} anticipated the problem and suggested a way to
overcome it: If both stars are directly observable, their relative proper
motion would (together with the light curve) supply $\theta_E$, while a second
observatory in the solar system would provide $r_E$.  Several related ideas
have been forwarded, notably by \cite{2002ApJ...572..521A}, who exploited a
combination of caustic-crossing times and finite source size effects to obtain
the first microlensing mass measurement.  A similar strategy has recently been
used by \cite{2009ApJ...698L.147G} to measure the mass of a brown dwarf to
10\%.

Another possibility is to measure the lensing effect of a star on a {\em
galaxy.} \cite{1996AcA....46..291P} pointed out that as a Galactic star makes a
near-transit of a galaxy, the latter will undergo a small shift in its apparent
centroid.  For this variety of microlensing to be observable, $\theta_E$ needs
to be larger than the precision of image centroiding--although it can be below
resolution--and proper motions need to be large.  Nearby stars are the only
realistic prospect, since
\begin{equation}
\theta_E \approx 0\bsec09
\left( \frac{M/M_\sun}{D_{\mathrm{lens}}/{\mathrm{pc}}} \right)^{1/2}
\label{eq:theta-to-mass}
\end{equation}
for sources at infinity, and proper motions are $\sim1''$ per year.
Microlensing by nearby stars would have none of the degeneracy problems
mentioned above; the lens being at known distance, the mass is the only unknown
parameter.  Moreover, such events are predictable long in advance.  The transit
may be on the order of a few months allowing for observations of the galaxy
both before and during the lensing event. Ideally, the star would not transit
directly across the galaxy since the star must be masked out for proper
observations of the galaxy. 

\cite{1996AcA....46..291P} suggested that microlensing centroid shifts could be
used to measure down to masses of nearby brown dwarfs.  In this paper, we
suggest a refinement of Paczy\'nski's idea which could make it much more
effective.  Rather than just the galaxy centroid shift, the whole weakly-lensed
image of the galaxy could be exploited to infer the lensing mass.  We develop a
technique to extract the weak-lensing effects and estimate the mass of the
star.

\section{A fitting method} \label{sec:Theory}

Consider a star with Einstein radius $\theta_E$ whose sky position at
epoch $t$ is $z_t$.  Lensing by this star maps a point $\theta$ in the image
plane to a point $\phi$ in the source plane such that
\begin{equation}
\label{eq:Lens equation}
\phi = \theta - \theta^2_E\frac{(\theta - z_t)}{|\theta-z_t|^2}
\end{equation}
where $z_t,\theta,\phi$ are two-dimensional vectors.

Next, we consider a galaxy whose unlensed brightness distribution $S$
is expanded in terms of basis functions $B^n(\theta)$ as
\begin{equation}
\label{eq:surface brightness}
S(\theta) = \sum_n a^n B^n(\theta)
\end{equation}
where $a^n$ are the expansion coefficients.  In the presence of
lensing, the brightness distribution of the galaxy will be
\begin{equation}
S_t(\theta) = \sum_n a^n B^n(\phi(\theta,z_t,\theta_E))\quad,
\end{equation}
where $S$ now has a time dependence due to the position of the star $z_t$.
Since lensing conserves surface brightness, the lensed surface brightness
at $\theta$ equals the unlensed surface brightness at $\phi$.

Pixelating the image plane, we write the pixel-wise brightness distribution as
\begin{equation}\label{modelsb}
D_{t,ij} \equiv \sum_n a^n L^n_{t,ij}
\end{equation}
where $L^n_{t,ij}$ represents a lensed and pixelated basis function
\begin{equation}
L^n_{t,ij} \equiv B^n(\phi(\theta_{ij},z_t,\theta_E))\quad.
\end{equation}
The expression in \eqnref{modelsb} is our model for the lensed brightness
distribution.  If the observed distribution is $d_{t,ij}$ then the likelihood
is
\begin{equation}
  \mathcal L(a^n,\theta_E) \propto \prod_{t,ij} \exp
  \big[ {\textstyle-\frac12} \, \sigma^{-2}_{t,ij} \,
        (d_{t,ij} - D_{t,ij})^2 \big]
\end{equation}
where $\sigma_{t,ij}$ is the pixel-wise noise, which we assume is Gaussian.
The model has a complicated dependence on $\theta_E$, but only a linear
dependence on the expansion coefficients $a^n$.  Since we are not especially
interested in the $a^n$, we marginalise them out by standard methods \citep[for
example, chapter~5 of][]{2003pda.book}.  The marginalised likelihood is given
by
\begin{equation}
2 \ln\mathcal L(\theta_E) = \ln |\det C|
+ \sum_{mn} P_mP_n C_{mn}
- \sum_{t,ij} \sigma^{-2}_{t,ij} \, d^2_{t,ij}\quad.
\label{eq:log likelihood}
\end{equation}
Here
\begin{equation}
P_n \equiv \sum_{t,ij} \sigma^{-2}_{t,ij} \, d_{t,ij} \, L^n_{t,ij}
\label{eq:data proj}
\end{equation}
represents a kind of projection of the data on the model, while
\begin{equation}
C^{-1}_{mn} \equiv \sum_{t,ij} \sigma^{-2}_{t,ij} \, L^m_{t,ij} \, L^n_{t,ij}
\label{eq:inv_covar}
\end{equation}
is the inverse covariance matrix.  We can relate $\mathcal L(\theta_E$) to an
effective $\chi^2$ as just 
\begin{equation}
\mathcal L(\theta_E) = \exp(-\chi^2/2)\quad.
\end{equation}

We are now prepared to estimate the mass of a star by its lensing effect on a
background galaxy. Given the pixel-wise brightness $d_{t,ij}$ and noise level
$\sigma_{t,ij}$, we simply need to calculate the effective $\chi^2$ and
minimize with respect to $\theta_E$. 

As an aside, $|\det C|$ in \eqnref{eq:log likelihood} will typically overflow
floating-point arithmetic, while $\ln |\det C|$ will fit quite
nicely. Hence we compute an alternative form, namely, $\ln |\det C| = \sum \ln
\lambda_{n}$, where $\lambda_{n}$ is the $n$th eigenvalue of $C$.

\section{Simulated Observations} \label{sec:Calculating Probabilities}

We tested the above scheme using simulated data, focusing specifically
on the dependence on four quantities: the actual Einstein radius of the
star $\theta_{E,\mathrm{true}}$, the closest approach or impact
parameter $p$, the number of observed epochs (or $t$ values)
$N_\mathrm{obs}$, and the total number of photons collected
$\gamma_\mathrm{tot}$.  The values are summarized in
\tabref{Table:Test parameters}. The full matrix of these parameters
was tested, giving a total of 500 simulated observational programs.
\begin{table}
\begin{center}
\begin{tabular}{ccc}
\hline
$\gamma_{\mathrm{tot}}$    & $\frac12$,1,3,7,12                             & $\times 10^6$ \\
$N_{\mathrm{obs}}$         & $2,3,4,5$                                      &               \\
$\theta_{E,\mathrm{true}}$ & $10, 20, 30, 40, 50$                           & [milliarcsec] \\
$p$                        & 0.08, 0.126, 0.232, 0.454, 0.903               & [arcsec]      \\
\hline
\end{tabular}
\end{center}
\caption{Parameters used for simulated observations:
  $\gamma_{\mathrm{tot}}$ is the total number of photons over all
  observation epochs (including one unlensed observation),
  $N_{\mathrm{obs}}$ is the number of epochs,
  $\theta_{E,\mathrm{true}}$ is the Einstein radius we wish to
  recover, and $p$ is the closest approach of the star on the sky
  plane to the background galaxy.}
\label{Table:Test parameters}
\end{table}

For the exact form of the unlensed surface brightness in
\eqnref{eq:surface brightness} we chose
\begin{equation}
S(\theta) = \exp\left(-7.67\left[\sqrt{(\theta/R_e)^2 + R_c^2}\right]^{1/4}\right)
\end{equation}
with $R_e=1$.  This is simply a de Vaucouleurs profile
modified by a core radius $R_c$ of 2~pixels to mimic the effect of the
telescope PSF on a singular cusp.

We then considered $71\times71$ pixels imaging a patch of sky $2''$ on a side
and centred on the galaxy.  With this resolution each pixel is about
$l=0\bsec028$ across, which, for example, would be equivalent to about one pixel
per resolution element of the Nasmyth Adaptive Optics System (NAOS)
Near-Infrared Imager and Spectrograph (CONICA) camera
\citep{2003SPIE.4841..944L,2003SPIE.4839..140R}.  At a redshift of $z\sim0.5$,
where we expect to find most of our background galaxies, $1''$ is about 6 kpc.
The corresponding lensed pixelated brightness is 
\begin{equation}
S_{t,ij} = S(\phi(\theta_{ij},z_t,\theta_{E,\mathrm{true}}))\quad.
\end{equation}
We normalized this brightness to the total number of photons
\begin{equation}
\sum_{t,ij} S_{t,ij} = \gamma_{\mathrm{tot}},
\end{equation}
and thus $\gamma_\mathrm{tot} / N_\mathrm{obs}$ is the number of photons per image.
Considering the normalized $S_{t,ij}$ as the expectation value of the pixel-wise
photon count, we then drew the simulated data $d_{t,ij}$ from a Poisson
distribution.  The noise level $\sigma_{t,ij}$ was taken as $\sqrt{d_{t,ij}}$,
which greatly simplifies Equations~\ref{eq:log likelihood} and \ref{eq:data proj}.

In practice, the star will be masked out during observations. To
simulate this, we discarded the pixels within one pixel-width of the star.

As basis functions we chose two-dimensional Hermite functions or shapelets
\citep[e.g.,][]{2003MNRAS.338...35R}.  The scale parameter for the
shapelets was set to $\beta=0\bsec2$ and we used
$N_\mathrm{basis}=20\times20$ basis functions in all.  These settings allow
structures as large as
$\theta_\mathrm{max}=\beta(2\sqrt{N_\mathrm{basis}}+1)^{1/2}=1\bsec28$ and as
small as
$\theta_\mathrm{min}=\beta(2\sqrt{N_\mathrm{basis}}+1)^{-1/2}=0\bsec03$ to be
resolved \citep{2007A&A...463.1215M}.

For each of the 500 test cases, we considered one epoch $t=0$ with the galaxy
unlensed and additional epochs $t>0$ with the star at $z_t=(5lt,{-l2^i\;|\;i
\in [1,\dots,5]})$. The selection of $i$ corresponds to the selection of $p$: A
large~$p$ implies a large~$i$. This choice of position puts the star at the
centre of a pixel. Extensive testing has shown a sensitivity to placing stars
near pixel edges, whereby recovery of the data is degraded if the star is too
close to pixel boundaries.

\begin{figure}
\plotone{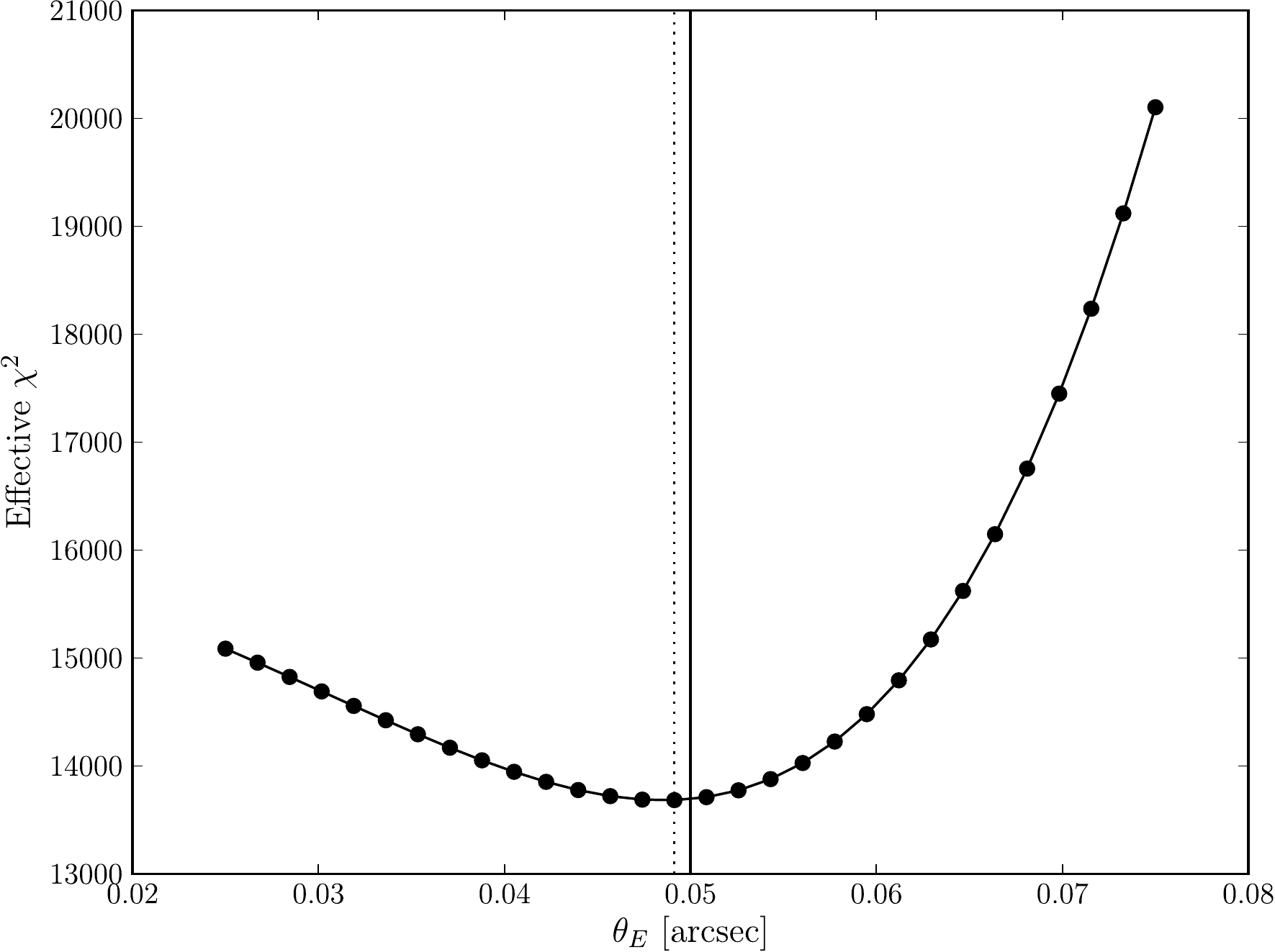}
\caption{Plot of the effective $\chi^2$ as a function of $\theta_E$,
for a simulated observation program with $N_\mathrm{obs}=2$ epochs,
total photons $\gamma_\mathrm{tot}=3\times10^6$ collected,
$\theta_{E,\mathrm{true}}=0\bsec05$ and impact parameter $p=0\bsec126$.
The solid vertical line marks $\theta_{E,\mathrm{true}}$ and the
dashed line marks the best fit.  The formal count of degrees of
freedom is $2\;\mathrm{epochs}\times71^2\;\mathrm{pixels}-20^2\;\mathrm{basis}\
\mathrm{functions}=9682$, so the effective reduced
$\chi^2\approx1.4$. The dependence of the fit on $\theta_E$ is
non-linear, hence the asymmetric shape of the curve.}
\label{fig:chi2}
\end{figure}

\begin{figure}
\plotone{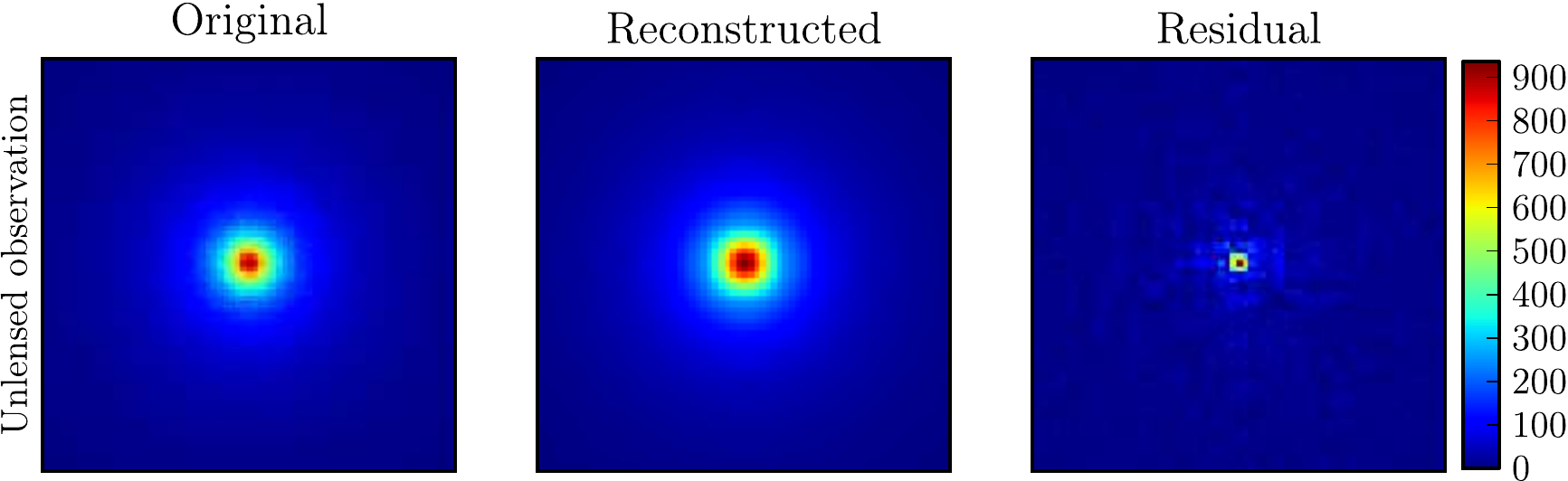}
\plotone{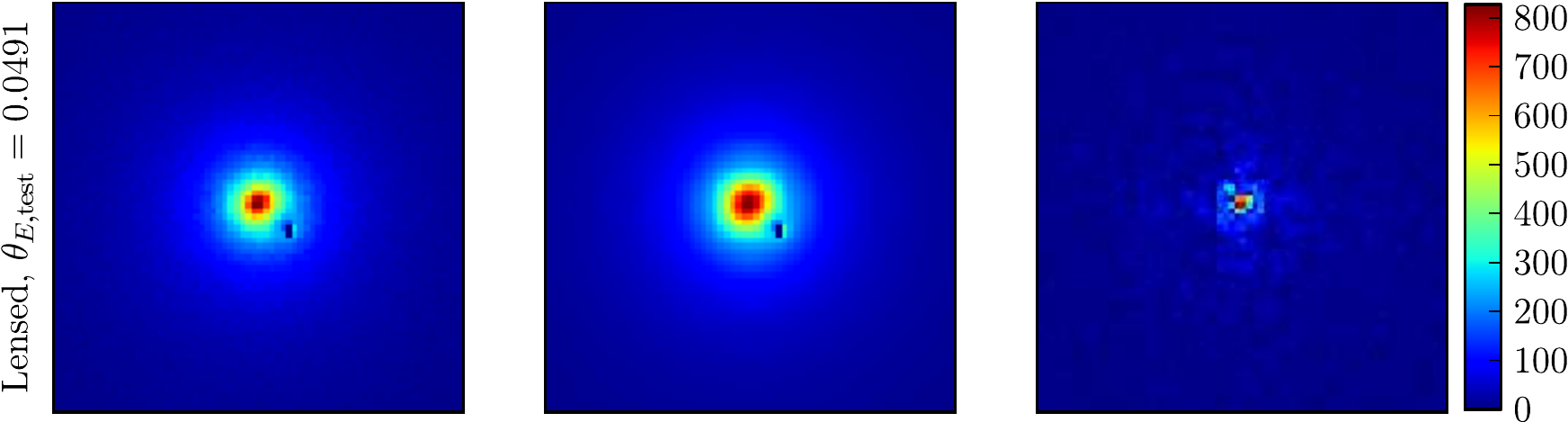}
\caption{Details of the best-fit $\theta_E$ for the simulated
observation program of \figref{fig:chi2}.  The upper row refers to an
unlensed epoch $t=0$ and the lower row to the lensed epoch $t=1$.  The
left column shows the simulated data $d_{t,ij}$, the middle column
shows the best reconstruction $D_{t,ij}$, while the right column is
$|d_{t,ij}-D_{t,ij}|$. The scale is in units of photons. The star has been
masked out as would be done during an observation.}
\label{fig:bestfit} 
\end{figure}

In \figref{fig:chi2} we show the effective $\chi^2$ as a function of $\theta_E$
for one of the simulated data sets. In \figref{fig:bestfit} we show the
simulated images, along with the reconstructed and residual images for the
best-fit $\theta_E$.  Examining such plots is a good indicator of potential
problems.  For example, if too few basis functions are used, a grid-like pattern
shows up in the reconstructions and the residual, and recovery of $\theta_E$ is
very poor.

\begin{figure}
\plotone{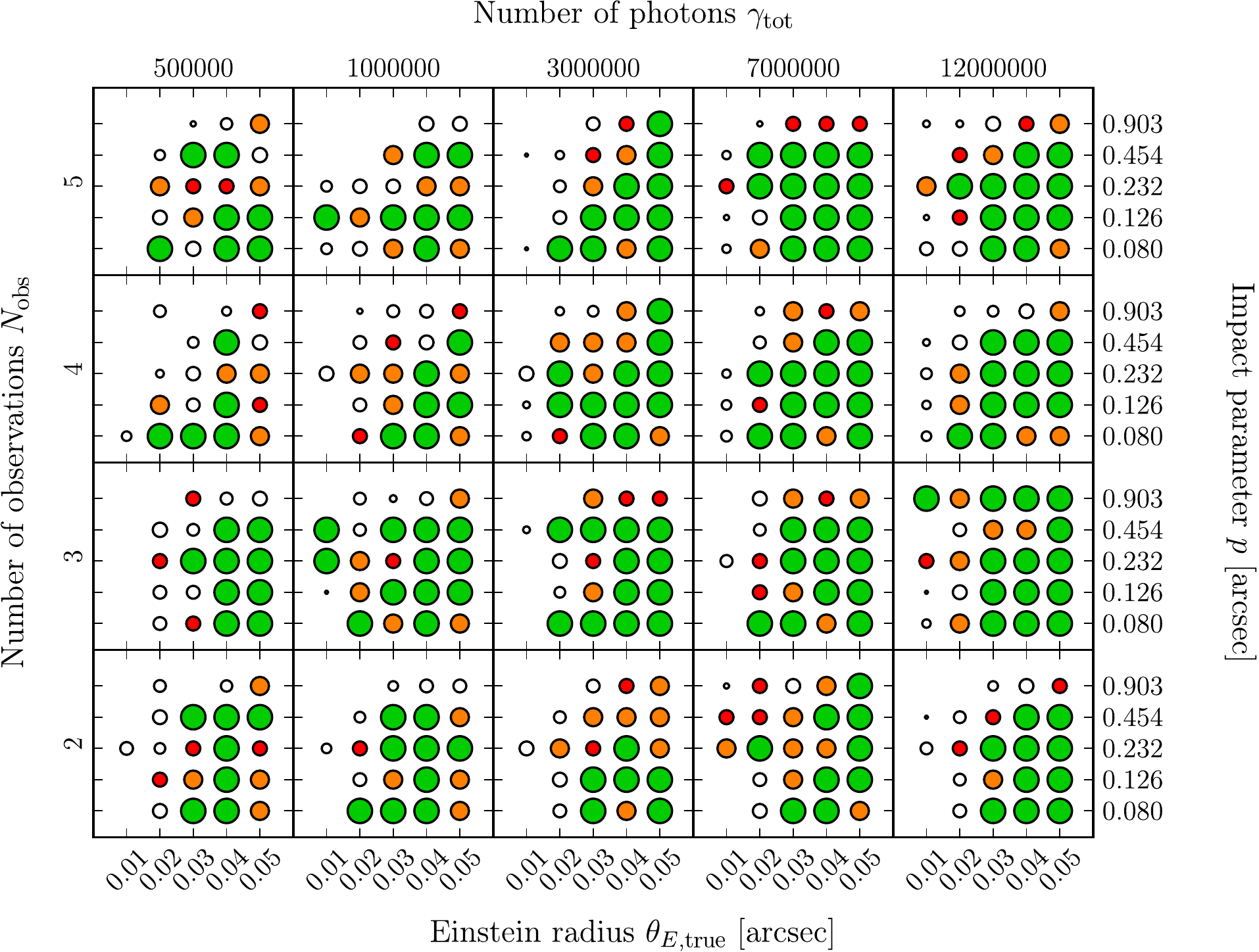}
\caption{Mass uncertainty as a function of \emph{four} quantities:
$\theta_E$, impact parameter $p$, total galaxy photons collected
$\gamma_\mathrm{tot}$, and the number of epochs $N_\mathrm{obs}$.
Within each box, $\theta_E$ and $p$ are varied at fixed
$\gamma_\mathrm{tot},N_\mathrm{obs}$.  The latter two quantities are
varied between boxes, as labelled.  Circles indicate the error in mass
(i.e., in $\theta_E^2$): Large, green circles denote $<5\%$ error,
orange circles 5--11\% error, red circles 11-20\% error, and small, open circles
are used if the mass error was $>20\%$.  Missing circles mean that no
mass estimate could be made. Of all the tests, 26\% have errors less than
5\% and 39\% have errors less than 11\%. Of those tests with filled circles,
52\% have errors less than 5\% and 79\% have errors less than 11\%.}
\label{fig:results}
\end{figure}

\figref{fig:results} summarizes the complete suite of 500 simulated observation
programs, showing the mass-recovery errors as a function of
$\theta_{E,\mathrm{true}},$ $p,$ $\gamma_{\mathrm{tot}}$ and
$N_{\mathrm{obs}}$. The following conclusions can be easily read off:
\begin{itemize}
\item The mass range of nearby brown dwarfs is accessible, since
  $\theta_E$ down to $0\bsec02$ can be measured with the 
  resolution considered.
\item Impact parameters of $p\sim10\,\theta_E$ are small enough, but if
  $p$ is too small the galaxy can be obscured by the star mask, leading
  to poor results.
\item Of order a million photons from the galaxy are needed, and a few
  times this are desirable, but it does not matter much whether these
  are concentrated in two epochs or distributed among several epochs.
\end{itemize}

\section{Event rates} \label{sec:Actual Observations}

\begin{figure}
\plottwo{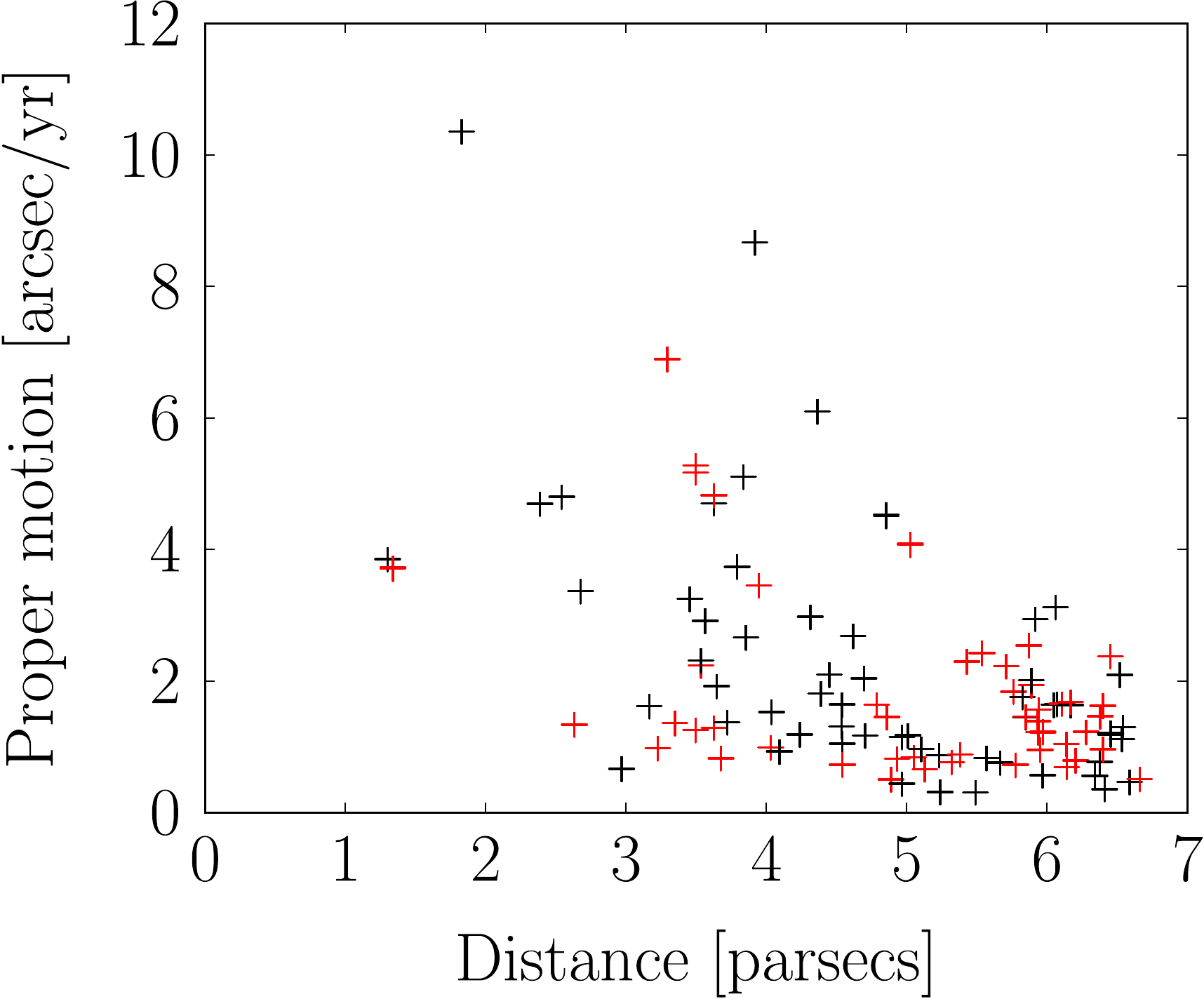}{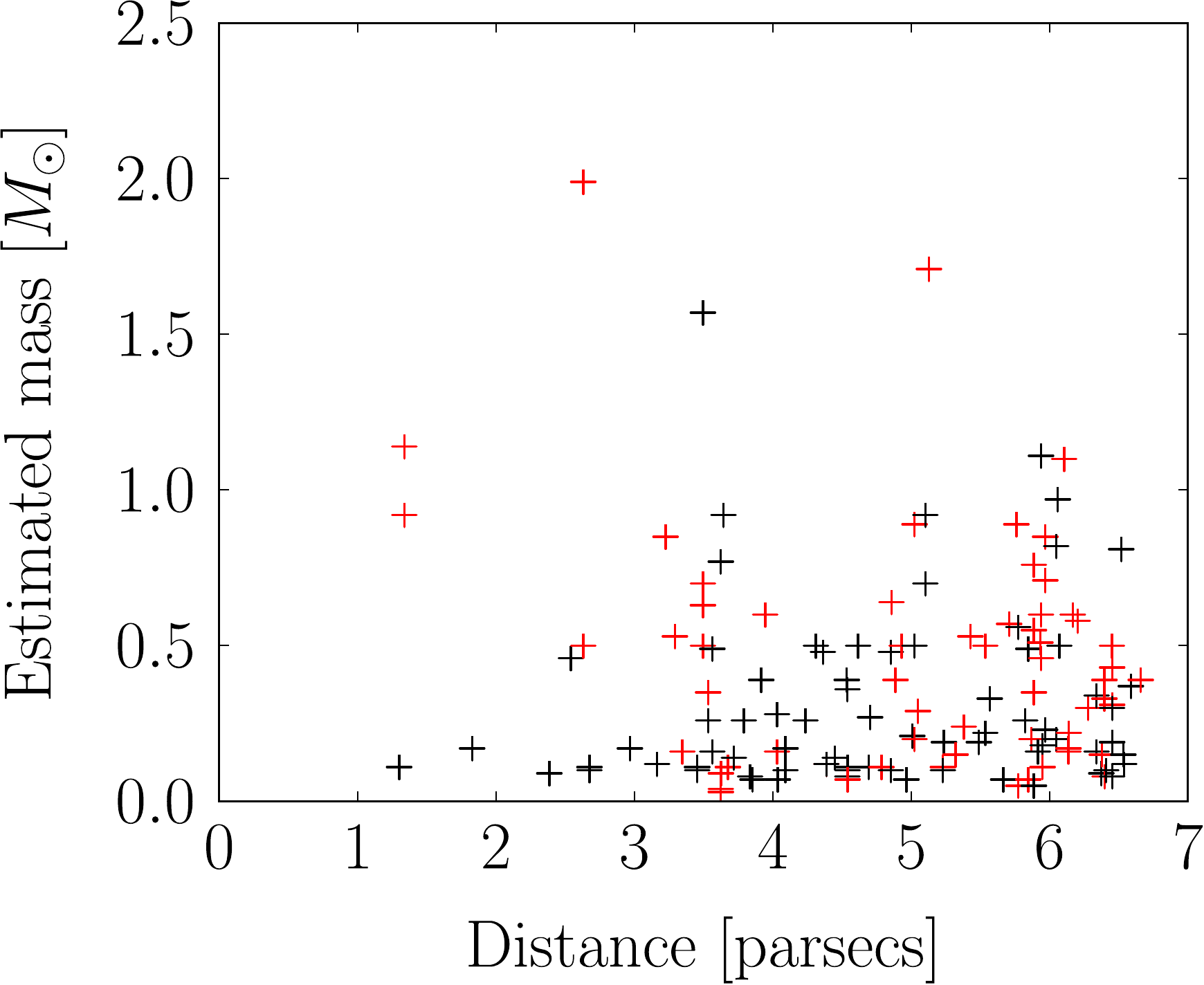}
\caption{The nearest 100 stellar systems (139 stars) from the RECONS catalogue.
Shown are the proper motions (left) and estimated masses (right) for the
sample.  Systems in red have been excluded from the discussion in
\secref{sec:Actual Observations}. The five highest mass stars are, $\alpha$
Centauri A+B, Sirius, Procyon, and Altair.}
\label{fig:RECONS data}
\end{figure}

\begin{figure}
\plottwo{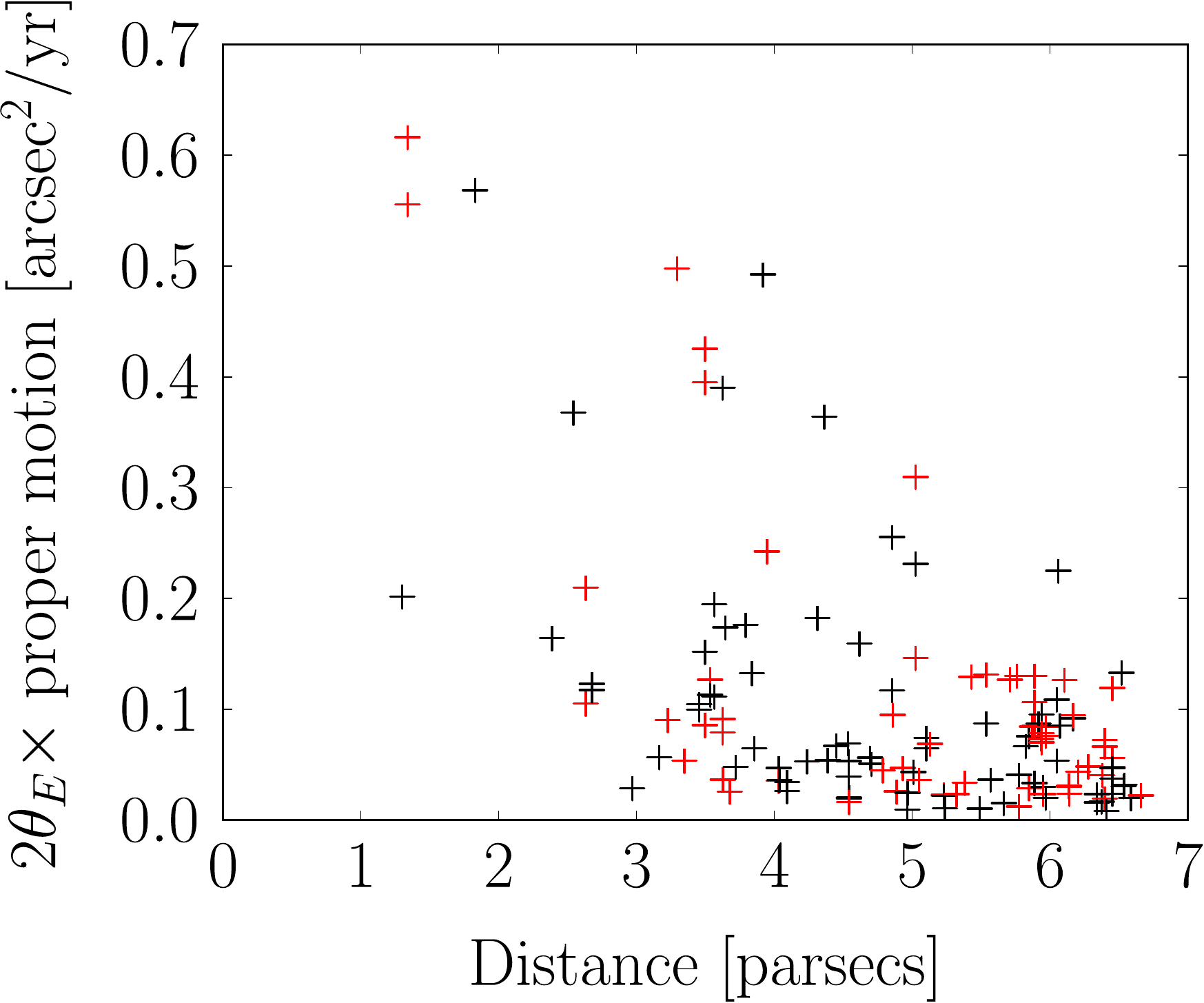}{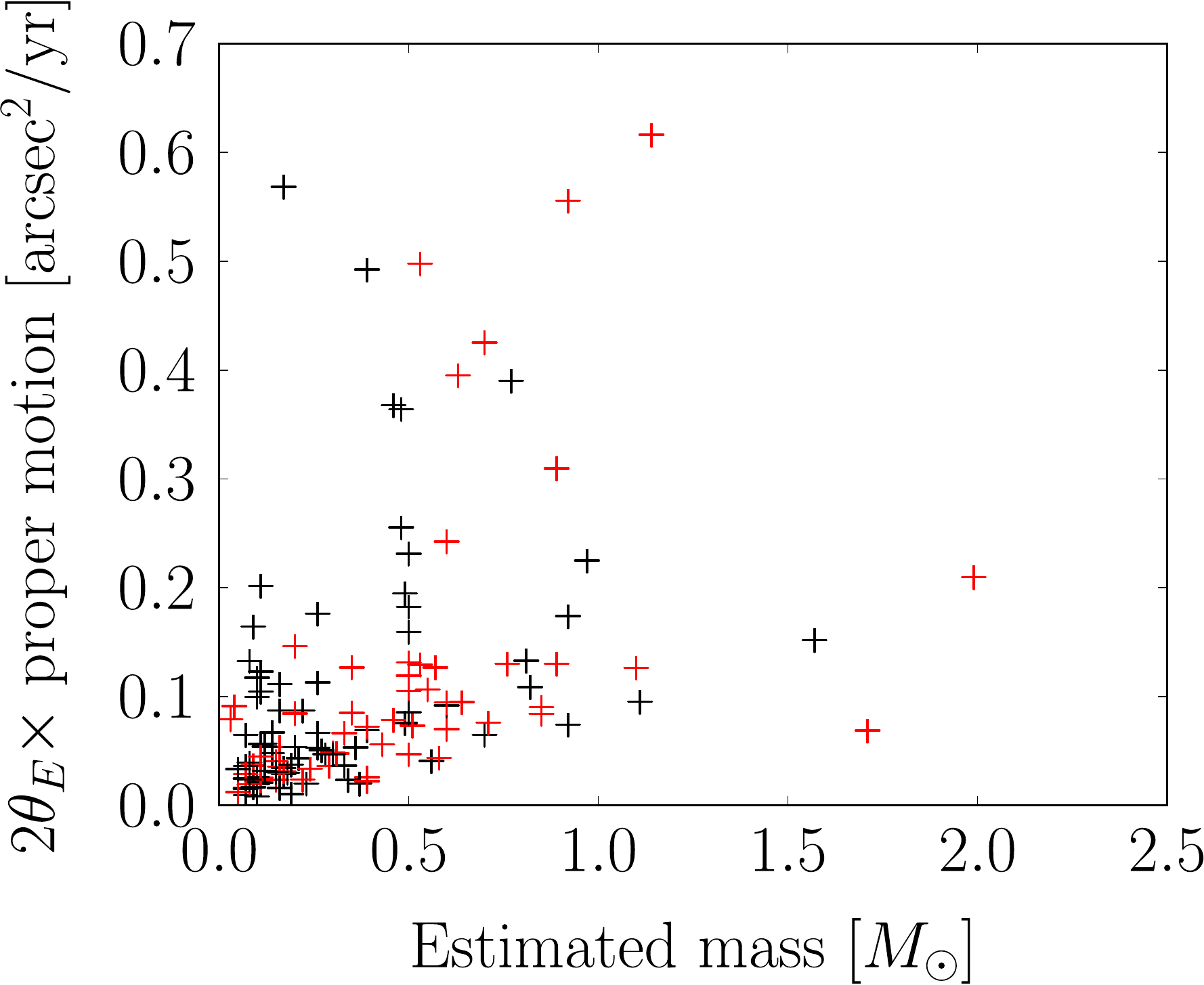}
\caption{Sky area coming within one Einstein radius of each nearby star per
year, plotted against distance (left), and against estimated stellar mass
(right). Systems in red have been excluded from the discussion in
\secref{sec:Actual Observations}.}
\label{fig:RECONS calc}
\end{figure}

We now consider how likely is it to find a star crossing near a background
galaxy.  For this analysis we used the Research Consortium on Nearby Stars
(RECONS) list of the 100 nearest stellar systems \citep{reconswebsite}.
The proper motions and estimated masses of the stars in these systems
are plotted in \figref{fig:RECONS data}. In \figref{fig:RECONS calc} we have
plotted the area of sky swept out by Einstein radii per year, or
$2\,\mu\theta_E$ where $\mu$ is the proper motion.

If we restrict ourselves to masses $<0.5 M_\sun$ and proper motions
$>0\bsec5\rm\,/yr$, we are left with 85 stars.  Assuming, as seen in our tests,
that a galaxy within $10\,\theta_E$ is a candidate, we sum $20\,\mu\theta_E$
over these stars.  The total available area is $\sim70\,\rm arcsec^2$ per year.

The GalaxyCount program \citep{2007MNRAS.377..815E} estimates
$\simeq1$ galaxy with magnitude $I\leq25$ within a sky area of $70\,\rm arcsec^2$.
This provides a rough estimate of the rate of observable weak
microlensing events.

\section{Observational Prospects}

Observing the weak lensing of a faint galaxy by a nearby star would require a
high resolution ($\approx 0\bsec05$) imager with high contrast capabilities. A
$0.5 M_\sun$ star at 5 pc has a brightness of $I\approx 6.5$mag while $I\approx
12$mag for a $0.1 M_\sun$ star at the same distance.  Thus a contrast in the
range $\Delta I =12-18$mag must be achieved at a separation of about 0\bsec05
for typical events.  This is quite a challenge for existing instruments.
Fortunately, rapid progress can be expected in this field by instruments
currently built for the imaging of planetary systems with 8m-10m telescopes and
further significant progress will be possible with extremely large telescopes
and high contrast imagers in space. They will provide very high contrast
observations $\Delta m > 20$mag and allow mass determinations of many nearby
stars using weak microlensing of faint background galaxies as advocated in this
Letter.  Any background light that is not from the galaxy can still be
considered part of the source as it will either be lensed or remain relatively
constant throughout the duration of the complete observation program. An 8m
class telescope with 30-50\% efficiency collects about 50,000 photons/hr. Thus,
a typical program might need between 20 and 100 hours to expect reasonable
results.

With existing instruments it should already be possible to observe weak
microlensing in favourable cases where the impact parameter is small
and the optical resolution is higher than that considered here. Nearby
($d\approx 5$ pc) brown dwarfs with a mass of $\approx 0.05 M_\sun$ ($\theta_E
\approx 0\bsec01$) such as SCR 1845-6357 B (at 3.9 pc), DENIS 0255-4700 (5.0
pc), 2MASS 0415-0935 (5.7 pc), or GJ 570 D (5.9 pc), have $I\approx 17-20$~mag
and $V\approx 22-25$~mag and they are not or not much brighter than the
abundant backgound galaxies. Low mass stars and substellar objects are red or
extremely red and imaging observations of blue star-forming galaxy at short
wavelengths is favoured because the image contamination of the lensed galaxy by
the PSF of the lensing object is strongly reduced.  It seems that HST or an
adaptive optics systems (e.g. with laser guide star) at a large telescope
working at short wavelengths $< 1\mu\rm m$ should be capable of achieving
successful observations for certain weak microlensing events.

\bibliographystyle{mn2e}
\bibliography{ms}

\label{lastpage}

\end{document}